\documentclass[journal]{IEEEtran}
\usepackage{cite}
\usepackage{graphicx}
\usepackage[colorlinks=true,urlcolor=blue,linkcolor=blue,citecolor=blue]{hyperref}
\usepackage{nidanfloat}
\ifCLASSOPTIONcompsoc
  \usepackage[caption=false,font=normalsize,labelfont=sf,textfont=sf]{subfig}
\else
  \usepackage[caption=false,font=footnotesize]{subfig}
\fi

\begin{document}
\title{Longitudinal Mode-by-Mode Feedback System for The J-PARC Main Ring}

%\author{First~A.~Author,~\IEEEmembership{Member,~IEEE,}
%        Second~B.~Author,~\IEEEmembership{Fellow,~OSA,}
%        and~Third~C.Author,~\IEEEmembership{Life~Fellow,~IEEE}% <-this % stops a space
%        
\author{Yasuyuki Sugiyama,
Fumihiko Tamura and
Masahito Yoshii%

%\thanks{Manuscript received November 4, 2007. (Write the date on which you submitted your paper for review.) This work was supported in part by the U.S. Department of Commerce under Grant No. BS123456 (sponsor acknowledgment goes here).}% <-this % stops a space'

\thanks{Manuscript received June 24, 2018.}% <-this % stops a space'

%\thanks{This research was supported by KEK, 
%MEXT KAKENHI Grant Number 18071006,
%JSPS KAKENHI Grant Number 23224007,
%the Japan/US Cooperation Program,
%the DOE award DE-SC0002644 through a subcontract from the University of Michigan,
%and DOE award DE-SC0006497 to Arizona State University.
%}%
\thanks{Yasuyuki Sugiyama, and Masahito Yoshii are with Accelerator Laboratory, High Energy Accelerator Research
Organization (KEK), 1-1 OHO, Tsukuba, Ibaraki 305-0801 Japan.
(e-mail: yasuyuki.sugiyama@kek.jp)
}%
\thanks{Fumihiko Tamura is with the J-PARC center, Japan Atomic Energy Agency (JAEA),
2-4 Shirakata, Tokai-Mura, Naka-Gun, Ibaraki, 319-1195 Japan.
(e-mail: fumihiko.tamura@j-parc.jp)
}%
%}%
%\thanks{Duncan McFarland and Joseph Comfort are with the Department of Physics, Arizona State Univeristy, Tempe, AZ 85287, USA.
%}%
%\thanks{Jiasen Ma and Yau Wai Wah are with the Enrico Fermi Institute, University of Chicago, 5640 South Ellis Ave, Chicago, IL 60637, USA. %(e-mail: jsma@uchicago.edu)
%}%
%
%\thanks{Yasuhisa Tajima is with the Department of Physics, Yamagata University, Yamagata, Yamagata 990-8560, Japan.% (e-mail: tajima@quark.kj.yamagata-u.ac.jp)
%}%
%\thanks{Mircea Bogdan is with the University of Chicago, 5640 South Ellis Ave, Chicago, IL 60637, USA.
%%(
%%telephone: 773-834-8483, 
%%e-mail: bogdan@edg.uchicago.edu). 
%}

%\thanks{Full names of authors are preferred in the author field, but are not required. Put a space between authors' initials. Do not use all uppercase for authors' surnames.}%
%\thanks{F. A. Author is with the National Institute of Standards and Technology, Boulder, CO 80303 USA (telephone: 303-497-3650, e-mail: author @boulder.nist.gov).}%
%\thanks{S. B. Author, Jr., was with Rice University, Houston, TX 77005 USA. He is now with the Department of Physics, Colorado State University, Ft. Collins, CO 80523 USA (telephone: 970-491-6206, e-mail: author@lamar. colostate.edu).}%
%\thanks{T. C. Author is with the Electrical Engineering Department, University of Colorado, Boulder, CO 80309 USA, on leave from the National Research Institute for Metals, Tsukuba, Japan (e-mail: author@nrim.go.jp).}%
}

\maketitle
\thispagestyle{empty}

%\begin{abstract}
%These instructions provide guidelines for preparing manuscripts for submission to the Conference Record (CR) of the 2014 IEEE Real Time Conference. If you are using {\LaTeX} to prepare your manuscript, you may use this document as a template. Define all symbols used in the abstract. Do not cite references in the abstract. 
%\end{abstract}
\begin{abstract}
%The abstract goes here.
The J-PARC Main Ring (MR) is a high intensity proton synchrotron, which accelerates protons from 3 GeV to 30 GeV. 
The MR delivers $2.6\times10^{14}$ protons per pulse, which corresponds to the beam power of 500 kW,
to the neutrino experiment as of May 2018,
and the studies to reach higher beam intensities are in progress. 
During  studies, we observed the longitudinal dipole coupled-bunch instabilities in the MR for the beam power beyond 470 kW.
To mitigate them for higher beam intensities, we have developed a longitudinal mode-by-mode feedback system.
The feedback system consists of a wall current monitor, a FPGA-based feedback processor, RF power amplifiers, and a RF cavity as a longitudinal kicker. 
In the feedback processor, we utilize the single sideband filtering technique to detect the oscillation components of the individual coupled-bunch mode in the beam signal.
%We present the design of the feedback system, especially the design detail of the digital filters in the feedback processor.
%We also report the preliminary beam measurement results for evaluation of the system performance on detection and excitation of the selected modes.
The frequency response of the filters in the feedback processor matched well with the simulation.
The oscillation amplitude of the coupled bunch oscillation measured by the system agreed with the oscilloscope analysis.

\end{abstract}


\section{Introduction}
%% no \IEEEPARstart
%This demo file is intended to serve as a ``starter file''
%for IEEE conference papers produced under \LaTeX\ using
%IEEEtran.cls version 1.8 and later.
%% You must have at least 2 lines in the paragraph with the drop letter
%% (should never be an issue)
%I wish you the best of success.
%\hfill December 27, 2012

%\subsection{particle and anti-particle}
%\subsection{CP Symmetry Breaking}
%	(Describe where your study stands in the field of physics, and why it is important.
%	Write it in a simple way so that first-year graduate students in other fields in physics can understand.)
%The material around us are made of atoms, and atoms are made of particles which are called as "proton", "neutron", and "electron".
%Electrons are  elementary particles. Protons and neutrons are not elementary particles, and they are made of  elementary particles called as "quarks".
%Electron and quark have their own partner particles which have similar characteristics except for opposite sign of electrical charge. Such partner particles are called as "anti-particles". 
%If we have large enough energy, we can generate a pair of particle and antiparticle from energy. If particles and anti-particles meet, they can turn into energy or photons.
%\IEEEPARstart{A}{ccording} to cosmology,
\subsection{J-PARC}
\IEEEPARstart{T}{he}
Main Ring synchrotron (MR)\cite{Koseki2012} in the
Japan Proton Accelerator Research Complex
(J-PARC)\cite{Nagamiya2012}
is a high intensity proton synchrotron which accelerates protons from 3 GeV to 30 GeV. 
% is a high intensity proton accelerator facility,
%which consists of the 400 MeV linac, %\cite{Ikegami2012,Hotchi:2014nka},
%the 3 GeV RapidCycling Synchrotron (RCS), %\cite{Hotchi2012,Hotchi:2014nka},
%and the 30 GeV Main Ring (MR)\cite{Koseki2012}.
The MR delivers the proton beams to the neutrino experiment 
%\cite{Sekiguchi2012} 
by the fast extraction (FX),
and to the hadron experiments
%\cite{Agari2012} 
by the slow extraction (SX).
%A schematic view of the MR is shown in Fig.\ref{fig:MR},
%and the 

\begin{figure}[hbtp]
\centering
\includegraphics[width=\linewidth]{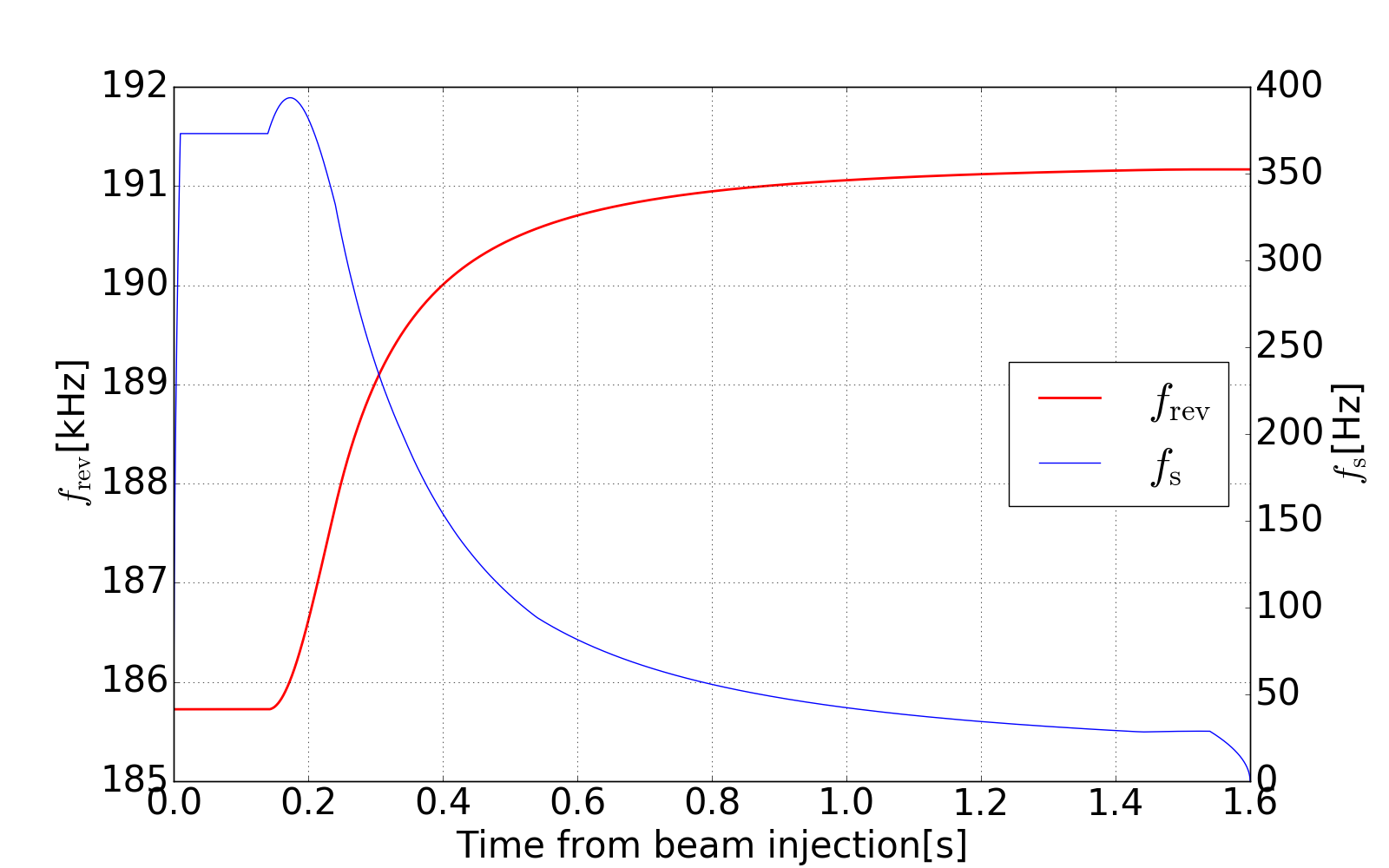}
\caption{The revolution frequency $f_\mathrm{rev}$ and the synchrotron frequency $f_\mathrm{s}$ in the J-PARC MR from the injection to the extraction.}
\label{fig:frev}
\end{figure}

The parameters of the MR and its RF system for the FX are shown in Table \ref{table:mrparam}.
Figure \ref{fig:frev}
% and \ref{fig:fs} 
 shows the  revolution frequency, $f_\mathrm{rev}$, and the synchrotron frequency, $f_\mathrm{s}$, in the MR. %from the injection to the extraction.
During the acceleration from 3 GeV to 30 GeV,
the synchrotron frequency is changing largely from 350 Hz at the injection to 30 Hz at the extraction along with the change of the revolution frequency from 185 kHz to 191 kHz.

%\begin{figure}[hbtp]
%\centering
%\includegraphics[width=\linewidth]{pts07101.png}
%\caption{The schematic view of the J-PARC MR.\cite{Koseki2012}}
%\label{fig:MR}
%\end{figure}

\begin{table}[hbt]
\centering
\caption{Parameters of the J-PARC MR and its RF system for the FX.}
\begin{tabular}{ll}
\hline
%\toprule
parameter&value\\
\hline
circumference& 1567.5 m\\
energy& 3--30 GeV\\
beam intensity& (achieved) $2.6\times10^{14}$ ppp\\
beam power& (achieved) 500 kW\\
repetition period& 2.48 s\\% (1.4 s for the acceleration)\\
accelerating period& 1.4 s\\
accelerating frequency $f_\mathrm{RF}$& 1.67--1.72 MHz\\
revolution frequency $f_\mathrm{rev}$& 185--191 kHz\\
harmonic number $h_\mathrm{RF}$& 9\\
number of bunches $N_b$& 8\\
maximum rf voltage& 320 kV\\
No. of cavities& 7 (h=9), 2 (h=18)\\
%second harmonic cavities& 2\\
Q-value of rf cavity& 22\\
\hline
%\bottomrule
\end{tabular}
\label{table:mrparam}
\end{table}

\begin{figure}[htbp]
\centering
\includegraphics[width=\linewidth]{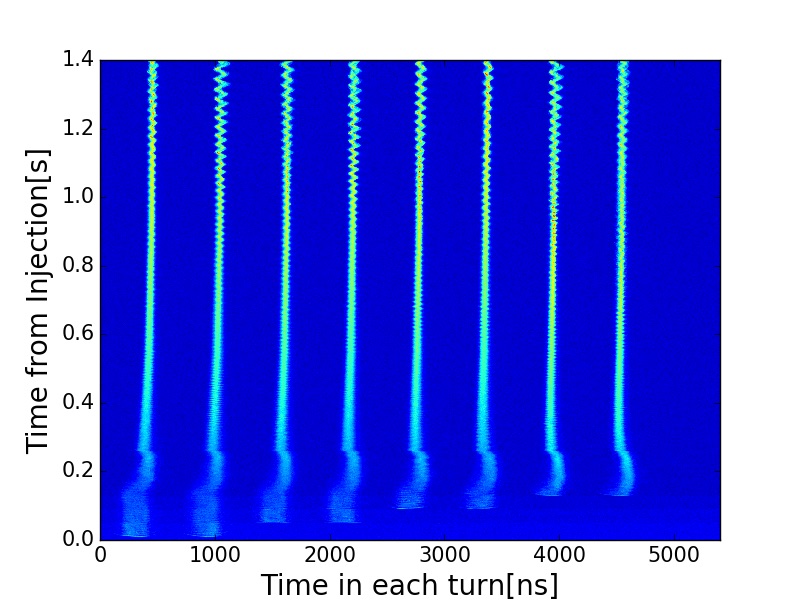}
\caption{The mountain plot for the fast extraction in the J-PARC MR with the beam power of 480 kW during the studies.\cite{Sugiyama:IBIC2017-TUPCF10}}
\label{fig:mountain}
\end{figure}

%\begin{figure}[hbt]
%\centering
%%\includegraphics[width=80mm]{mountain.jpg}
%\includegraphics[width=\linewidth]{TUPCF10f1.jpg}
%\caption{The synchrotron frequency in the J-PARC MR from the injection to the extraction}
%\label{fig:fs}
%\end{figure}
%

The MR delivers $2.6\times10^{14}$ protons per pulse, which corresponds to the beam power of 500 kW,
to the neutrino experiment as of May 2018.
Large longitudinal bunch oscillation was observed in the MR for the beam power beyond 470 kW\cite{Sugiyama:IBIC2017-TUPCF10},
%,
%and studies toward higher beam intensity are in progress.
%During studies, the longitudinal bunch oscillation 
%and appeared to be an issue to achieve higher beam intensities than 500 kW.
and it is necessary to suppress the oscillation for stable beam acceleration at the beam power higher than 500 kW.
\subsection{Longitudinal Bunch Oscillation in the J-PARC MR}
The longitudinal beam oscillation can be seen in a mountain plot. %which shows the change of the longitudinal profile.
Figure \ref{fig:mountain} shows the typical mountain plot 
of the beam signal at the beam power of 480 kW. % for the fast extraction during the studies. 
The beam signal from an Wall Current Monitor (WCM)\cite{toyama} is recorded by an oscilloscope. %,
%LeCroy WP715Zi, with the sampling frequency of 500 MHz.
The longitudinal dipole oscillation keeps growing from the middle of the acceleration.
%and their amplitudes keep increasing until the extraction. 
Each bunch has different amplitude and phase of the dipole oscillation.
%One can notice that the amplitudes and the phases of the oscillations of 
%the bunches are different.
This kind of the oscillations is
called the coupled bunch (CB) oscillation,
and the instability caused by the CB oscillation is called the CB instability (CBI).
%The identification of the CB oscillation mode
%% and corresponding frequency
% is necessary to find the source of the CBI.

\subsection{Coupled Bunch Oscillation}
\begin{figure}[t]
\centering
\includegraphics[width=\linewidth]{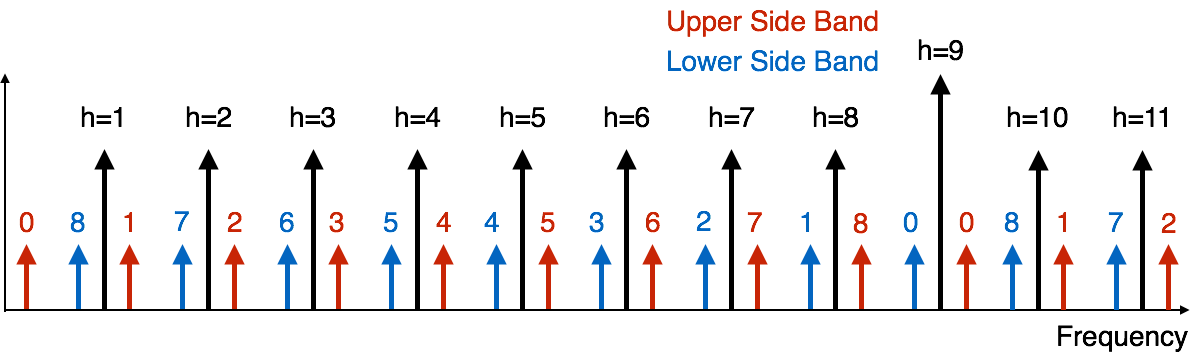}
\caption{The spectra of the synchrotron sidebands corresponding to the coupled bunch oscillation mode for the J-PARC MR\cite{Sugiyama:IBIC2017-TUPCF10}.}
\label{fig:HandModes}
\end{figure}
For M bunches, there are M modes of the CB oscillation with the mode number $n=0 ... M-1$.
%The phase difference of the synchrotron oscillation between adjacent bunches is $2\pi n/M$.
%For each mode,
%all bunches oscillate with the same frequency and the amplitude but with different phases.
The CB modes can 
%also 
be seen in the spectrum of the beam signal as synchrotron sidebands of the harmonic components\cite{Pedersen1977}.
The frequency of the CB mode $n$ can be expressed as follows:
\begin{equation}
f_{p,m,n}=|(pM+n)f_\mathrm{rev}+mf_s|,(-\infty<p<\infty)
\label{f_mode},
\end{equation}
where %$n$ represents the CB mode,
$f_\mathrm{rev}$ is the revolution frequency,
$f_s$  the synchrotron frequency,
and $m$  the type of the synchrotron motion.
The case with $m=1$ corresponds to the dipole oscillation, and the case with $m=2$ corresponds to the quadruple oscillation.
%In this paper,
%we only focus on the dipole oscillation ($m=1$) which can be seen significantly in the MR.
%The CB modes appear as the Upper Side Band (USB) in the case with $p\geq0$,
%and as the Lower Side Band (LSB) in the case of $p<0$.
%%In the case with $m=1$,
The CB modes appear as the Upper synchrotron Side Bands (USBs) and
the Lower synchrotron Side Bands (LSBs) in the cases of $p\geq0$ and $p<0$, respectively. 
%For the CB mode $n$,
Below the accelerating frequency,
the  LSB and the USB with the CB mode $n$ can be expressed as follows:
\begin{eqnarray}
f_n^{\mathrm{USB}}&=&nf_\mathrm{rev}+mf_s\\
f_n^{\mathrm{LSB}}&=&(M-n)f_\mathrm{rev}-mf_s.
\label{fn}
\end{eqnarray}

There are 9 CB modes for the MR since the harmonic number of the MR is 9.
%For the MR, 
%$M=9$ and there are 9 CB modes since the harmonic number of the RF system $h_\mathrm{RF}=9$,
%though the maximum number of bunch for the operation is limited up to 8.
%there are CB modes 
The spectra of the synchrotron sidebands corresponding to the CB modes in the MR up to the harmonic $h=11$ are illustrated in Fig. \ref{fig:HandModes}.
% shows the CB modes below and around the acceleration harmonic $h=9$ for the J-PARC MR.
There are two sidebands with different CB modes in each harmonic component.
%The USBs closer to the acceleration harmonics $(h=9)$ have larger CB mode number,
%while the LSBs closer to the acceleration harmonics $(h=9)$ have smaller CB mode number.

Based on the analysis of the longitudinal CB oscillation
\cite{Sugiyama:IBIC2017-TUPCF10},
strong CB oscillation of mode $n=8$ was observed in the harmonic component of $h=8,10$.
%The sideband amplitude can be obtained by applying the single sideband filtering method\cite{Kriegbaum1977} to the beam signal.
%\begin{figure*}[b]
%\centering
%%\includegraphics[width=80mm]{usbmode.png}
%%\includegraphics[width=80mm]{usb1to11.png}
%%\includegraphics[width=\linewidth]{TUPCF10f8.png}
%\includegraphics[width=.8\linewidth]{oscilloana_480kW.pdf}
%\caption{The time variation of the amplitudes of the LSBs(left) and USBs (right) of the harmonic components\cite{Sugiyama:IBIC2017-TUPCF10}.}
%\label{fig:SSBresult_USB}
%\end{figure*}
%Figure \ref{fig:SSBresult_USB} % and \ref{fig:SSBresult_LSB}
% shows the time variation of the amplitudes of the USBs and LSBs of the harmonic components based on the sideband analysis\cite{Sugiyama:IBIC2017-TUPCF10} .
%Strong CB oscillation of mode $n=8$ can be seen in the harmonic component of $h=8,10$.
%\begin{figure}[hbtp]
%\centering
%%\includegraphics[width=80mm]{lsbmode.png}
%%\includegraphics[width=80mm]{lsb1to11.png}
%\includegraphics[width=\linewidth]{TUPCF10f9.png}
%\caption{The time variation of the amplitudes of the LSBs of the harmonic components\cite{Sugiyama:IBIC2017-TUPCF10}.}
%\label{fig:SSBresult_LSB}
%\end{figure}
The suppression of these CB oscillation is a key to achieve the beam power higher than 500 kW.

%\newpage
\section{Longitudinal Mode-by-Mode Feedback System}
%\subsection{system}
To mitigate CB instabilities, we have developed a longitudinal mode-by-mode feedback system.

\begin{figure}[t]
\centering
\includegraphics[width=\linewidth]{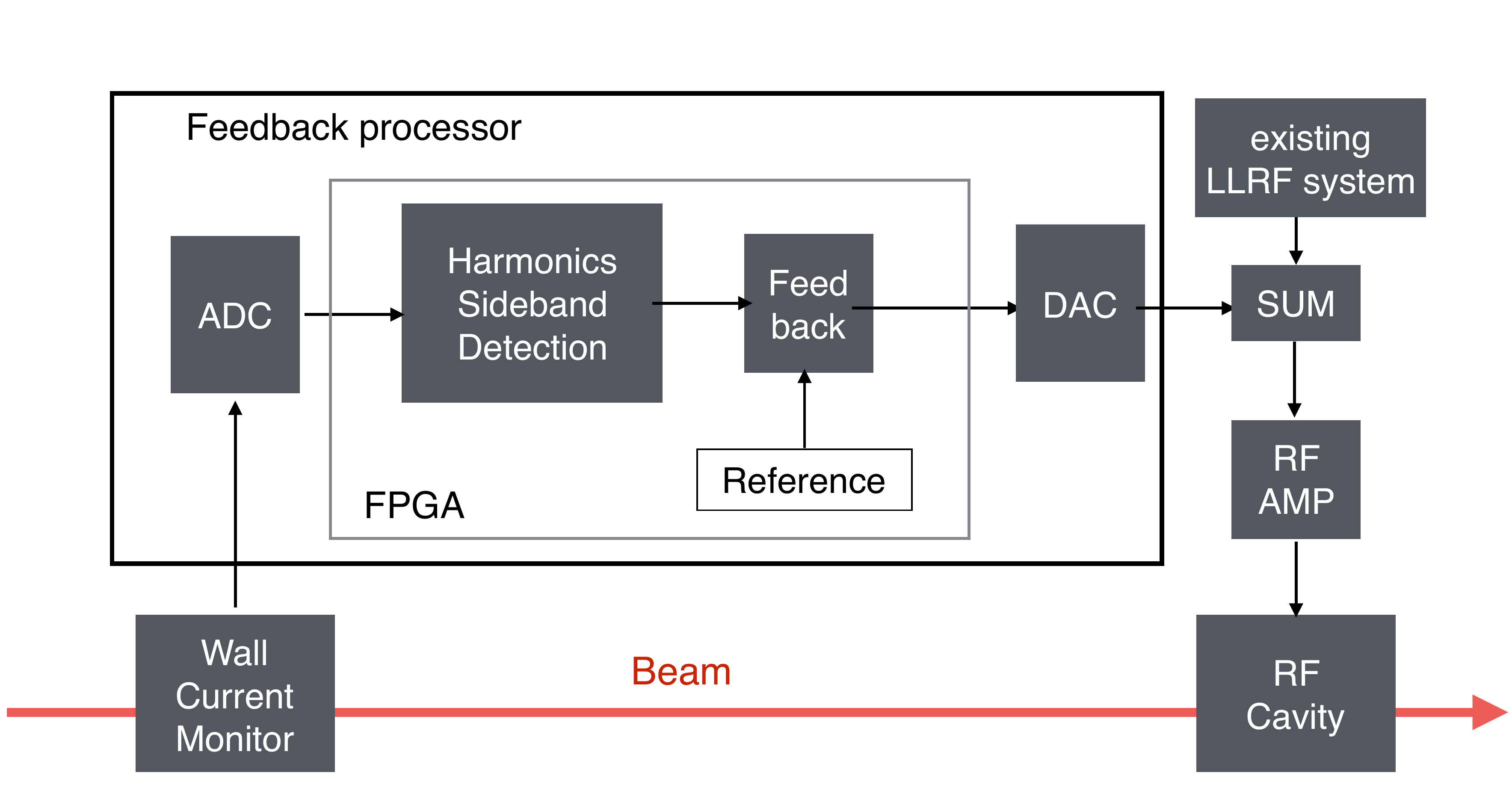}
\caption{Block diagram of the longitudinal mode-by-mode feedback system.}
\label{fig:FBsystem}
\end{figure}
\begin{figure}[t]
\centering
\includegraphics[width=.9\linewidth]{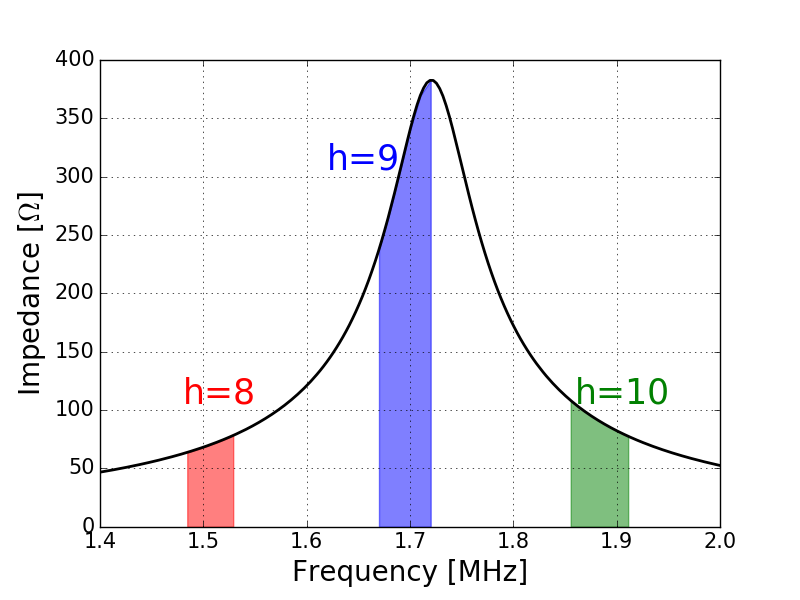}
\caption{The frequency response of the impedance of the RF cavity. }
\label{fig:impedance}
\end{figure}
Figure~\ref{fig:FBsystem} shows the block diagram of the longitudinal mode-by-mode feedback system.
The feedback system consists of a WCM, a FPGA-based feedback processor, RF power amplifiers, and a RF cavity as a longitudinal kicker. 
The beam signal is detected by a WCM and fed to a feedback processor.
The feedback processor detects the CB oscillation components of the beam signal and uses it for the feedback control.
The detail design of the feedback processor is discussed in the following section.
The feedback signal from the feedback processor is led to a high level RF (HLRF) system consisting of power amplifiers and a RF cavity.

The feedback system utilize the existing HLRF system used for the acceleration in the MR.
Figure \ref{fig:impedance} shows the frequency response of the impedance of the RF cavity used for the acceleration in the MR.
The RF cavity has impedance large enough to generate the kick voltage for the feedback in the frequency range for $h=8,10$ component,
and can be used as a longitudinal kicker.

The feedback signal from the feedback processor is summed with the RF signal from the existing low level RF (LLRF) system for the acceleration\cite{Tamura2013}.
The summed signal is amplified by the RF power amplifiers and fed into the 
 RF cavity not only to accelerate the beam but also to control the beam oscillation.
% for both of acceleration of the beam and suppression of the beam oscillation.

\section{Longitudinal Mode-by-Mode Feedback processor}

\begin{figure}[t]
\centering
\includegraphics[width=.95\linewidth]{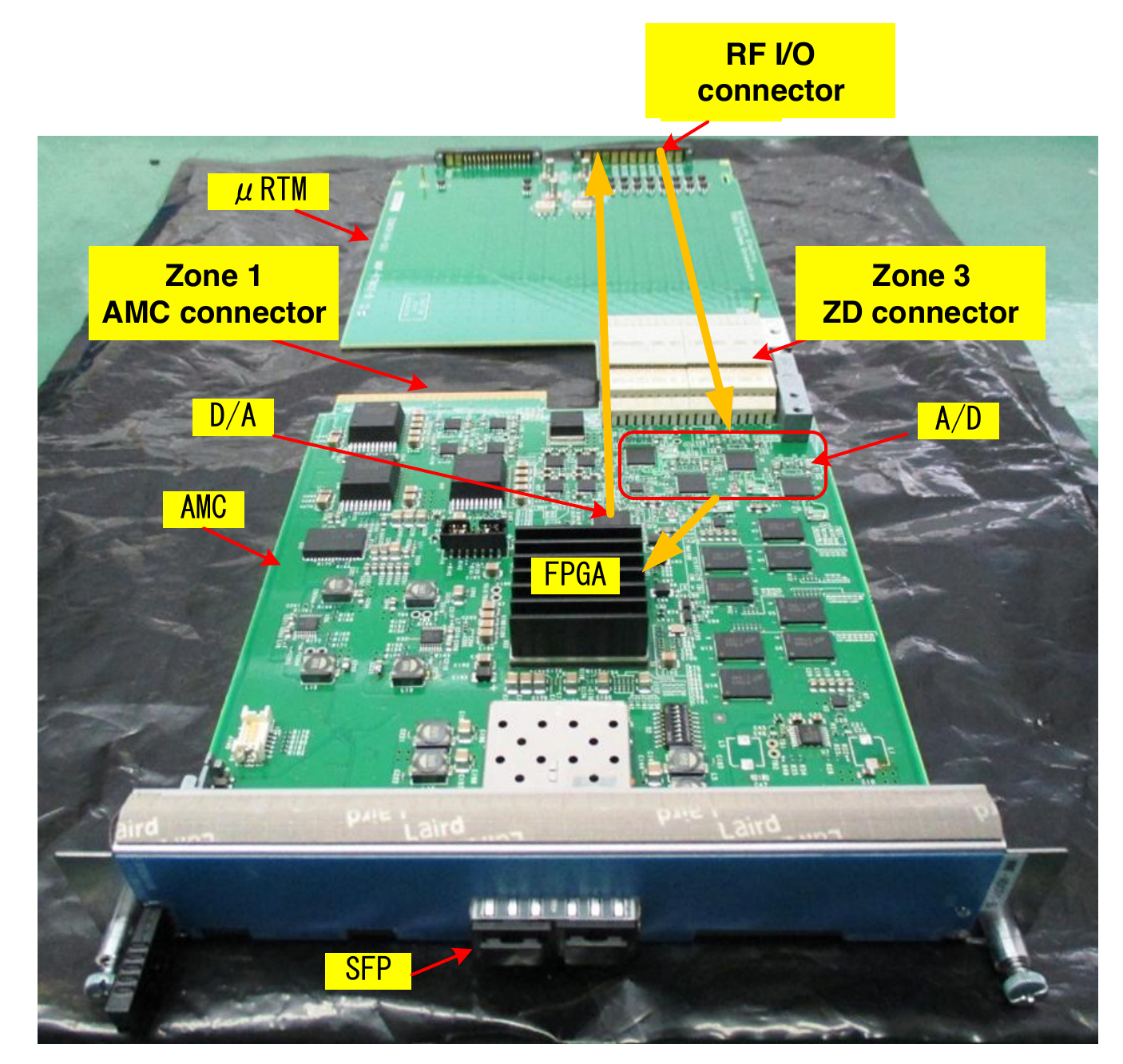}
\caption{Pictures of the longitudinal mode-by-mode feedback processor.\cite{Ryoshi2015}}
\label{fig:FBprocessor_photo}
\end{figure}

\begin{figure*}[b]
\centering
\includegraphics[width=\linewidth]{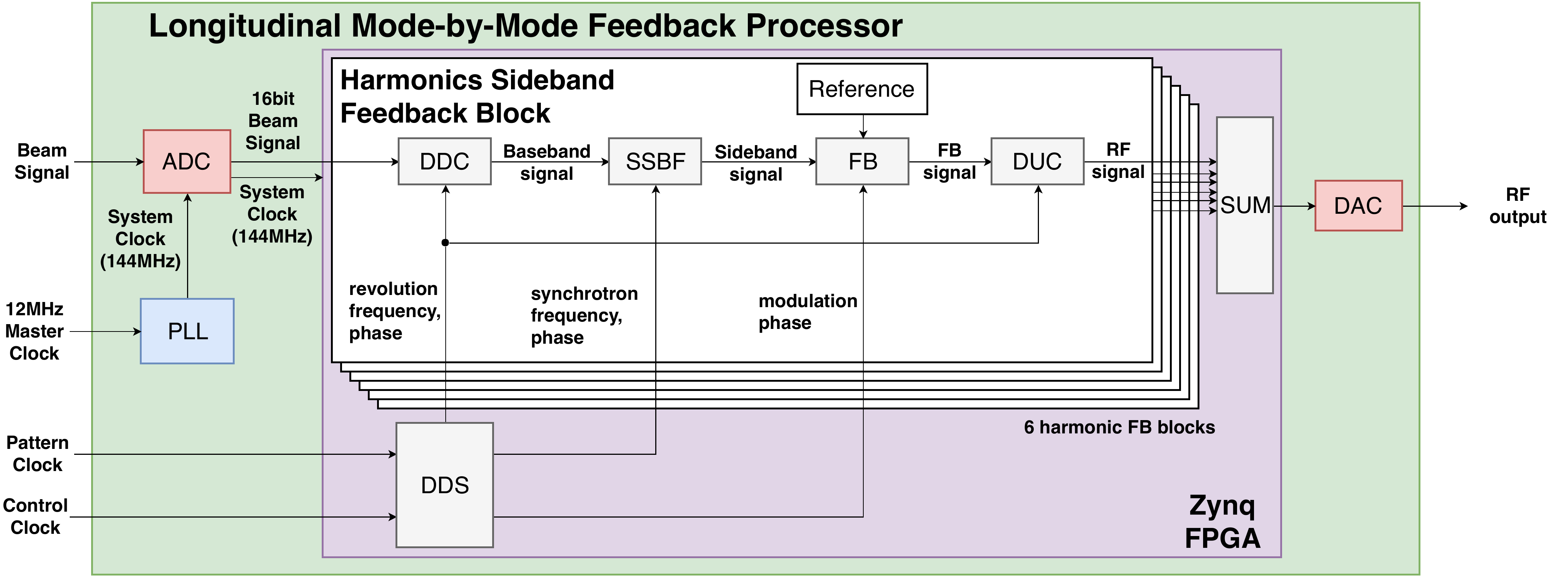}
\caption{Block diagram of the longitudinal mode-by-mode feedback processor.}
\label{fig:FBprocessor}
\end{figure*}

%\subsection{Requirement}
We developed a FPGA-based longitudinal mode-by-mode feedback processor for the feedback system.
The feedback processor was manufactured by Mitsubishi Electric TOKKI Systems Corporation based on the MicroTCA.4 architecture.

Figure \ref{fig:FBprocessor_photo} shows the pictures of the longitudinal mode-by-mode feedback processor.
The feedback processor consists of an AMC and a Rear Transition Module (RTM).

The general purpose AMC module\cite{Ryoshi2015} developed by Mitsubishi Electric TOKKI Systems Corporation is used in the system.
The AMC has 8 ADC channel by 4 ADC\footnote{Texas Instruments ADC16DX370 16-bit 370-MSPS ADC} chips, 2 DAC channel with a DAC\footnote{Analog Devices AD9783 16-bit 500-MSPS DAC} chip, and a FPGA.
Xilinx Zynq XC7Z045 SoC FPGA is used as a FPGA not only for the signal processing but also for the management of the module itself via EPICS IOC running on the embedded Linux.
The 1-GB DDR3-SDRAM on the AMC is used as a pattern memory to store the pattern. % such as the frequency and the gain.
A single slot AMC backplane is connected to the AMC to provide the ethernet connection and the power. % without the MicroTCA crate.
%An EPICS-IOC running on embedded Linux on the Zync enables a remote management of the module.

The RTM has analog and digital I/O ports and is used as the signal transition module for the AMC.
The RTM generates the 144-MHz clock signal for the ADC and the FPGA from the 12-MHz J-PARC master clock by a phase lock loop.

Figure~\ref{fig:FBprocessor} shows the block diagram of the longitudinal mode-by-mode feedback processor.
The feedback processor works at a system clock of 144 MHz from the RTM module.
The frequency and the phase signals are generated in the Direct Digital Synthesis (DDS) block.
The digitized waveform of the beam signal is converted into the baseband signal of each harmonic component by the Digital Down Converter (DDC).
%The synchrotron sideband components are detected by using the single sideband filter (SSBF)\cite{Kriegbaum1977},
and used for the feedback control.
The control of each CB mode is achieved by the feedback control of each synchrotron sideband component filtered by the single sideband filter (SSBF)\cite{Kriegbaum1977}.
The outputs of the feedback control are converted to the RF signal by the Digital Up Converter (DUC).
The RF signals are summed and converted to the analog signal by the DAC.
The feedback processor can process 6 different harmonic components at the same time.
The waveforms of the I/Q waveform at each processing stage were recorded and can be acquired via EPICS.

\subsection{Direct Digital Synthesis}
%
%\begin{figure}[t]
%\centering
%\includegraphics[width=\linewidth]{DDS.pdf}
%\caption{Block diagram of the Direct Digital Synthesis (DDS) block.}
%\label{fig:DDS}
%\end{figure}
The phase signals used for the signal demodulation and the modulation in the feedback processor are generated by the DDS.
The 34-bit phase accumulator generates the phase signal from the frequency signal.
The frequency patterns
% for the revolution frequency, the synchrotron frequency and the frequency for the sideband modulation
 are stored as 32bit words in the DDR3 memory,
and loaded from the memory in each pattern clock of 5 kHz.
%For the synchrotron frequency and the modulation frequency,
%the frequency information are led to the phase accumulator and the phase signals are generated.
%%The synchrotron frequency signals are generated from these frequency information and distributed to the SSBF.
%
The frequency pattern for the revolution frequency contains the initial frequency as a header word and latter words contains frequency offsets information.
The revolution frequency signal is generated by the frequency accumulator
which adds the frequency offset to the initial frequency in each control clock of 250 kHz.
%The revolution frequency signal is led to the phase accumulator and the revolution phase signal is generated.

\subsection{Digital Down Converter}
The digitized beam signal is down-converted into the baseband I/Q signals of each harmonic component by the DDC.

\begin{figure}[t]
\includegraphics[width=\linewidth]{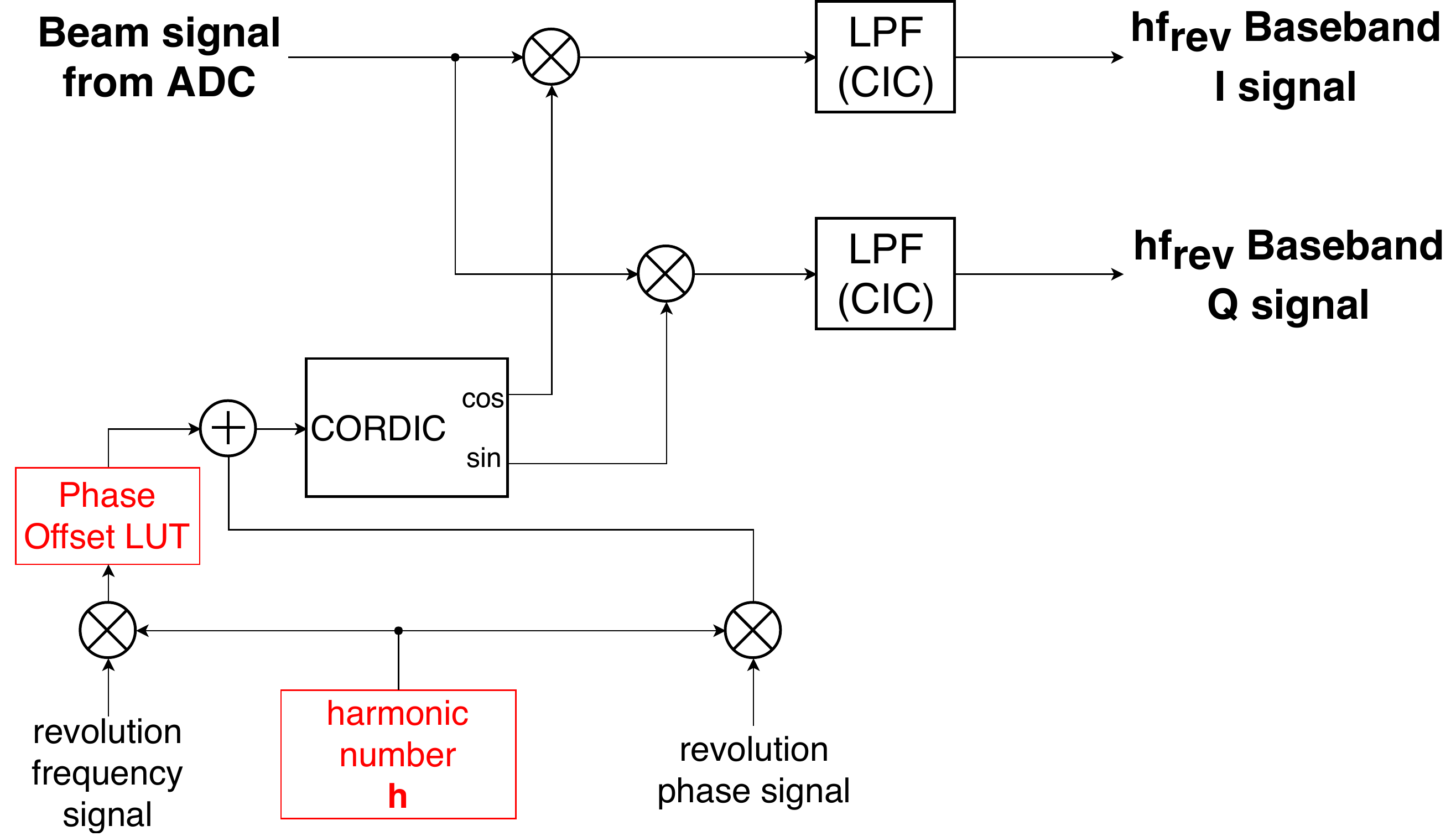}
\caption{Block diagram of the Digital Down Converter(DDC).}
\label{fig:DDC}
\end{figure}

Figure \ref{fig:DDC} shows the block diagram of the DDC.
The digitized beam signal is mixed with the sine and cosine signals of the harmonic frequency.
The sine and cosine signals are generated by the CORDIC using the phase signal of the harmonic.
%The revolution frequency pattern is set via EPICS. 
The phase signal of the harmonics is generated by multiplying the revolution phase signal with the selected harmonic number.
The phase offset to the CORDIC can be set  to compensate the phase frequency response of the system.
The LUT for the phase offset is addressed by the harmonic frequency. % and can be set via EPICS.

The mixed signals are led to the Low Pass Filter (LPF) and the baseband I/Q signal is detected.
The 5-stage CIC (Cascaded Integrator and Comb) filter is used as a LPF.
The CIC filter is running at 144 MHz as the sampling clock.
Its decimation ratio is 2 and the differential delay is 256 clocks.
%with decimation ratio of 2, differential delay of 256 clocks and 144MHz system clock as the sampling clock is used as LPF.
%The baseband I/Q signal can be detected from the outputs of LPF.

%\newpage
\subsection{Single Sideband Filter}
%To control the oscillation of each CB mode,
The LSB and USB
% of the synchrotron sidebands
% in the harmonic component 
 are detected separately by the SSBF. % and led to the feedback block.
%The feedback system is required to control and suppress only the oscillating component of the beam signal which is the sideband signal in the baseband harmonics signal.
%The suppression of the DC component and the transmission for the sideband signal in the harmonic baseband signal are key for the filter before the feedback block.
%To achieve this, the single sideband filter (SSBF) is implemented before the feedback block.

%The SSBF consists of the combination of the single sideband demodulator and modulator.

%\begin{figure}[t]
%\centering
%\includegraphics[width=.8\linewidth]{ssbf_principle.pdf}
%\caption{The principle of the Single Sideband filtering method.}
%\label{fig:SSBfilter_principle}
%\end{figure}

\begin{figure*}[b]
\includegraphics[width=\linewidth]{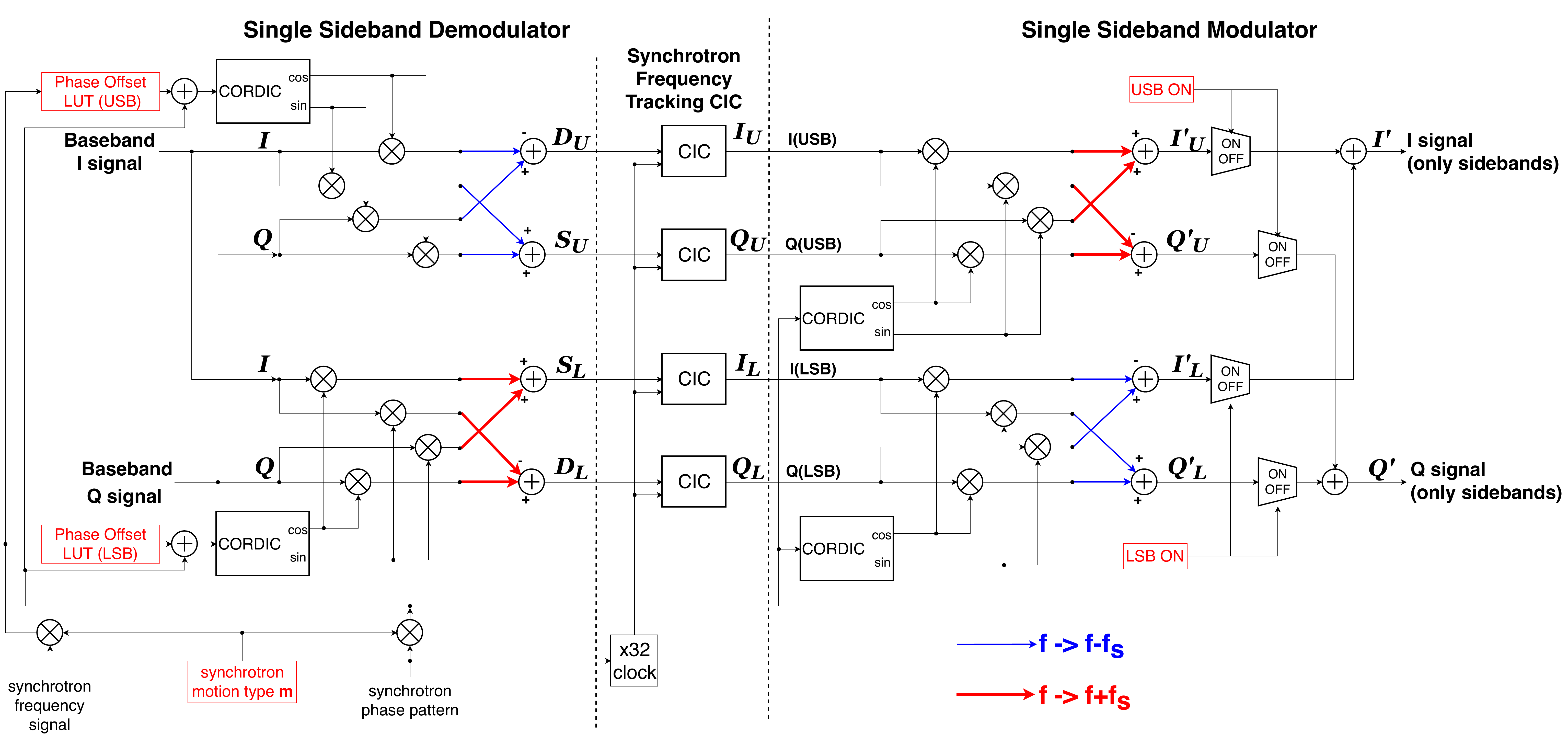}
\caption{Block diagram of the Single Sideband filter block.}
\label{fig:SSBfilter}
\end{figure*}

%Figure \ref{fig:SSBfilter_principle} shows the principle of the single sideband filtering method,
%and 
Figure \ref{fig:SSBfilter} shows the block diagram of the SSBF implemented in the feedback processor.

%In the baseband I/Q signal of the harmonic component,
%the synchrotron sidebands located at $f=\pm mf_s$. %, where the harmonic number $m$ corresponds to
%By mixing the baseband I/Q signals with the sine and cosine signal of $f=\pm mf_s$,
%the spectra are shifted so that the components of $f=\pm mf_s$ locate at $f=0$ after summation and subtraction.
%At the same time, $f=0,\mp mf_s$ components are shifted to the sideband with $f=\mp mf_s, 2\mp mf_s$, respectively.
%The LSB and the USB can be demodulated by suppressing these unwanted sidebands.
%After the demodulation, the baseband signal of the synchrotron sidebands are up-converted to the baseband of the harmonic component by a single sideband modulator. 
%After the down-conversion at the DUC,
The I/Q signals of the the harmonics component which contain the synchrotron sidebands are expressed as
\begin{eqnarray}
I&=&B\sin{\phi_B}+ \sum_{l=1..2} L_l\sin\left(-l\omega_s t +\phi_{L_l}\right) \nonumber\\
 & &+\sum_{l=1..2} U_l\sin\left(l\omega_s t +\phi_{U_l}\right)       \label{eqn:Il}\\
Q&=&B\cos{\phi_B}+ \sum_{l=1..2} L_l\cos\left(-l\omega_s t +\phi_{L_l}\right) \nonumber\\
& &+\sum_{l=1..2} U_l\cos\left(l\omega_s t +\phi_{U_l}\right) , \label{eqn:Ql}
\end{eqnarray}
where
% $m$ is the harmonic number for the synchrotron frequency,
$\omega_s$ is the synchrotron frequency satisfying $\omega_s=2\pi f_s$,
$L_l(\phi_{L_l})$, $B(\phi_B)$ and $U_l(\phi_{U_l})$ \ the amplitude (phase) of LSB, baseband, and USB components,
respectively,
and $l$ the type of the synchrotron motion.

The I,Q signals are mixed with $\sin m\omega_st$, $\cos m\omega_st$,
and summed or subtracted to get $S_U$,$D_U$,$S_L$,$D_L$:
\begin{eqnarray}
S_U&=&Q\cos\omega_st +I\sin\omega_st\nonumber\\
   &=&\sum_{l=1..2} L_l\cos\left(-l\omega_st-m\omega_st+\phi_{L_l} \right)\nonumber\\
   & & + B\cos\left(-m\omega_st+\phi_B \right)\nonumber\\
   & &+\sum_{l=1..2}U_l\cos\left(l\omega_st-m\omega_st+\phi_{U_l}\right)\\
%   \end{eqnarray}
%\begin{eqnarray}
D_U&=&I\cos\omega_st -Q\sin\omega_st\nonumber\\
   &=&\sum_{l=1..2} L_l\cos\left(-l\omega_st-m\omega_st+\phi_{L_l} \right)\nonumber\\
   & & + B\sin\left(-m\omega_st+\phi_B \right)\nonumber\\
   & &+\sum_{l=1..2}U_l\cos\left(l\omega_st-m\omega_st+\phi_{U_l}\right)\\%  
%\end{eqnarray}
%
%\begin{eqnarray}
S_L&=&I\cos\omega_st +Q\sin\omega_st\nonumber\\
   &=&\sum_{l=1..2} L_l\sin\left(-l\omega_st+m\omega_st+\phi_{L_l} \right)\nonumber\\
   & & + B\sin\left(m\omega_st+\phi_B \right)\nonumber\\
   & &+\sum_{l=1..2}U_l\sin\left(l\omega_st+m\omega_st+\phi_{U_l}\right)\\
%\end{eqnarray}
%\begin{eqnarray}
D_L&=&Q\cos\omega_st -I\sin\omega_st\nonumber\\
   &=&\sum_{l=1..2} L_l\cos\left(-l\omega_st+m\omega_st+\phi_{L_l} \right)\nonumber\\
   & & + B\cos\left(m\omega_st+\phi_B \right)\nonumber\\
   & &+\sum_{l=1..2}U_l\cos\left(l\omega_st+m\omega_st+\phi_{U_l}\right).%\\
 \end{eqnarray}
The harmonic number $m$ for the synchrotron frequency can be changed in the feedback processor to select the type of the longitudinal motion.
%set so that detecting component can be selected between the dipole and the quadruple oscillation component.
The sine and the cosine signals for mixing are generated by the CORDIC.
The phase offset to the CORDIC can be set separately for the LSB and USB to compensate the phase frequency response of the system.
The LUT for the phase offset is addressed by the harmonics of the synchrotron frequency. %and  can be set via EPICS.

The I/Q signal of USB(LSB), $I_U(I_L)$, $Q_U(Q_L)$ for the selected type of the synchrotron motion are obtained by applying a narrow LPF to $D_U(S_L)$,$S_U(D_L)$,
respectively:
\begin{eqnarray}
I_{U}&=&U_m\sin\phi_{U_m}\\
%   \end{eqnarray}
%\begin{eqnarray}
Q_{U}&=&U_m\cos\phi_{U_m}\\
%   \end{eqnarray}
%\begin{eqnarray}
I_{L}&=&L_m\sin\phi_{L_m}\\
%   \end{eqnarray}
%\begin{eqnarray}
Q_{L}&=&L_m\cos\phi_{L_m}.
\end{eqnarray}

The narrow LPF is required to suppress the sideband component at the harmonics of the synchrotron frequency which changes during the acceleration.
In the case of general CIC filter as a LPF,
its fixed frequency response changes its performance to suppress these sidebands along with the change of the synchrotron frequency.
This means the amount of the remaining sideband component is changed during the acceleration and makes it difficult to control only the selected sideband.
To achieve the suppression of these unwanted sidebands for the whole beam cycle, %which change their frequency during the acceleration,
the two-stage frequency tracking CIC filter is used as a narrow LPF.
%Figure \ref{fig:trackCICfilter} shows the block diagram of the frequency tracking CIC filter block.
%\begin{figure}[htbp]
%\includegraphics[width=\linewidth]{impedance.png}
%\caption{Block diagram of the frequency tracking CIC filter block.}
%\label{fig:trackCICfilter}
%\end{figure}
The frequency tracking CIC filter is the CIC filter which changes its notch position along with the frequency pattern.
This functionality is achieved by running the CIC filter with 32nd-harmonic of the 1st notch frequency and set its differential delay as 32 clocks.
With this setting, the CIC filter has notches at the harmonics of the 1st notch frequency.
By setting the synchrotron frequency pattern as the 1st notch frequency of the filter,
the filter can always suppress the unwanted sidebands while they change their positions during the acceleration.
%The 2-stage frequency tracking CIC filter is implemented in the feedback processor.

\begin{figure}[t]
\centering
\includegraphics[width=.8\linewidth]{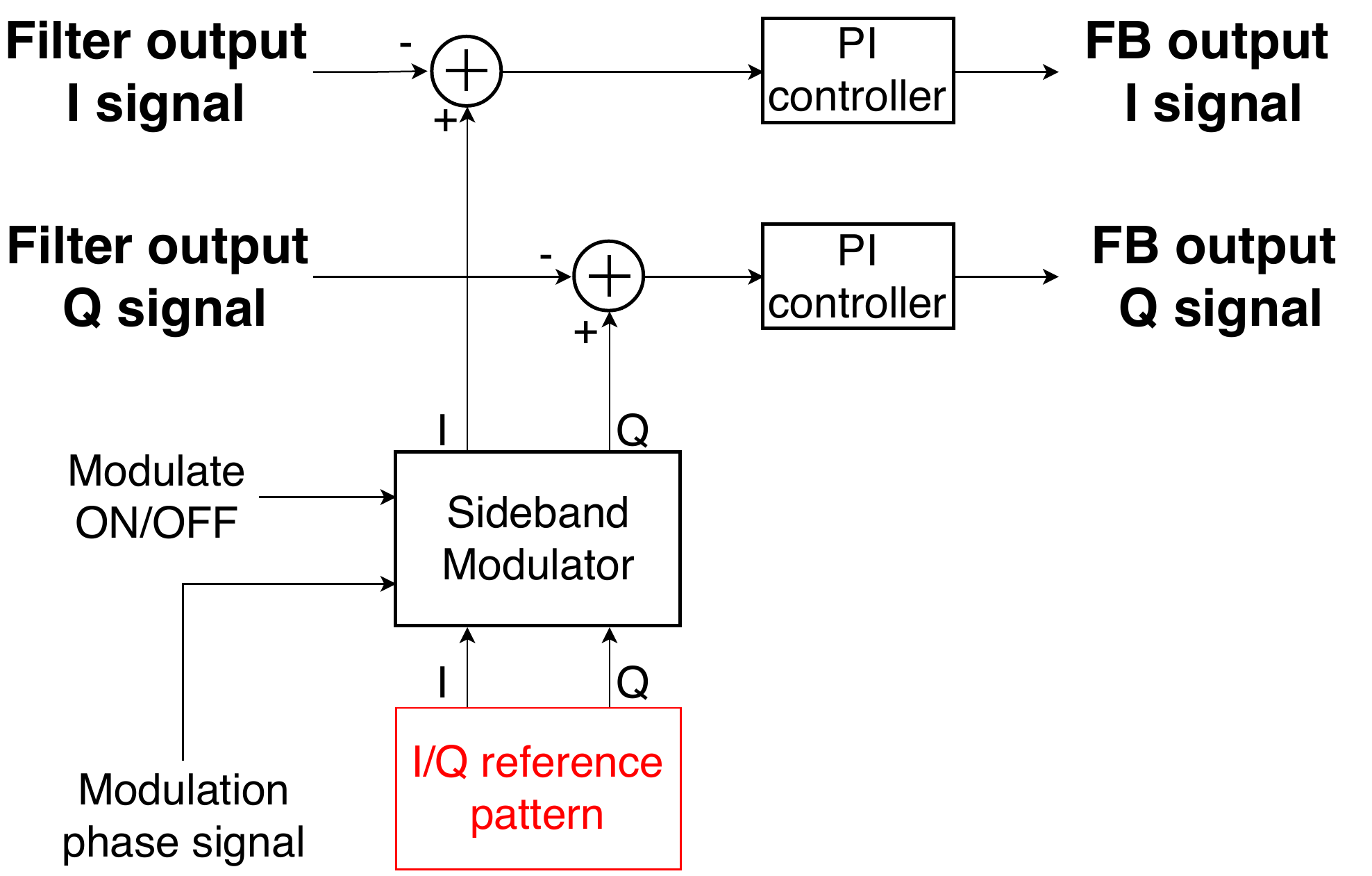}
\caption{Block diagram of the feedback block.}
\label{fig:FBblock}
\end{figure}

\begin{figure*}[b]
\centering
\includegraphics[width=.9\linewidth]{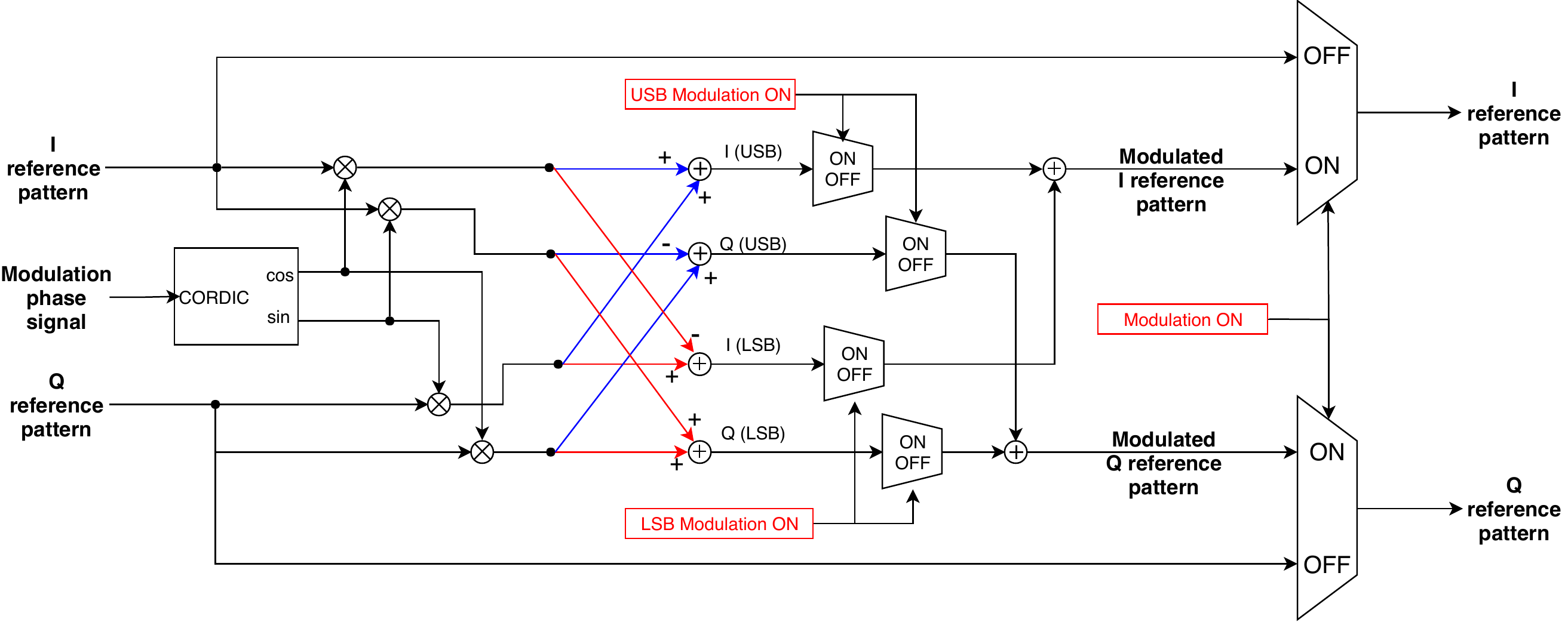}
\caption{Block diagram of the reference pattern modulation block.}
\label{fig:Modblock}
\end{figure*}

The I/Q signals of the  LSB and the USB are up-converted to the baseband signal of the harmonic component by mixing with the sinusoidal signal of the synchrotron frequency:
\begin{eqnarray}
I'_{U}&=&I_{U}\cos m\omega_st+Q_{U}\sin m\omega_st\\
Q'_{U}&=&Q_U\cos m\omega_st-I_U\sin m\omega_st\\
I'_{L}&=&I_{U}\cos m\omega_st-Q_{L}\sin m\omega_st\\
Q'_{L}&=&Q_L\cos m\omega_st+I_L\sin m\omega_st,
\end{eqnarray}
where $I'_{U}(I'_{L})$ and $Q'_{U}(Q'_{L})$ are the up-converted I and Q signals of USB(LSB) components, respectively.
After up-conversion, I/Q signal of the LSB and the USB are summed as the I/Q signal to the feedback block:
\begin{eqnarray}
I'&=&I'_{U}+I'_{L}\\
Q'&=&Q'_{U}+Q'_{L},
\end{eqnarray}
where $I'$ and $Q'$ are the baseband I/Q signals the harmonic components only with the sideband component.
The summation of the LSB and the USB can be disabled individually so that sidebands to be controlled can be selected.

%The frequency pattern of $f_\mathrm{rev}$ and $f_{s}$ are used for these filters.
%In the feedback processor, we utilize the single sideband filtering technique to detect the oscillation components of the individual coupled-bunch mode in the beam signal.
%The feedback processor can control the USB and LSB of 6 harmonic components simultaneously.

%\subparagraph[tracking filter]
%\subparagraph{CIC filter}
%
%\paragraph{High Pass Filter}
%As an alternative solution for the sideband filtering,
%a HPF is implemented in parallel with the SSBF.
%The HPF consists of an 1-stage comb filter and a Leaky Integrator (LI).
%The comb filter runs with 5kHz and its delay is 2 clock.
%The LI runs with 62.5kHz and  m=1024.

%\newpage

\subsection{Feedback Control Block}

Figure \ref{fig:FBblock} shows the block diagram of the feedback block.

For the feedback control, a Proportional (P) and an Integral (I) controller is implemented in the feedback logic.
The P controller runs with the system clock of the 144 MHz and the I controller runs with 250 kHz.
%The reference I/Q pattern used for the feedback can be set via EPICS.
The I/Q reference values are loaded from the pattern memory in each pattern clock.

The reference pattern can be modulated with the synchrotron frequency for the beam excitation measurement.
Figure \ref{fig:Modblock} shows the block diagram of the modulation block.
The reference I/Q pattern is mixed with the sine and the cosine signal of the modulation frequency as same way as that used in the single sideband modulator in the SSBF.
%The modulation frequency pattern can be set separately from that used for the sideband detection.
%Both the LSB and the  USB can be excited by the modulation and the sideband to be excited can be selected.
The sideband to be excited by the modulation can be selected.

\begin{figure}[t]
\centering
\includegraphics[width=\linewidth]{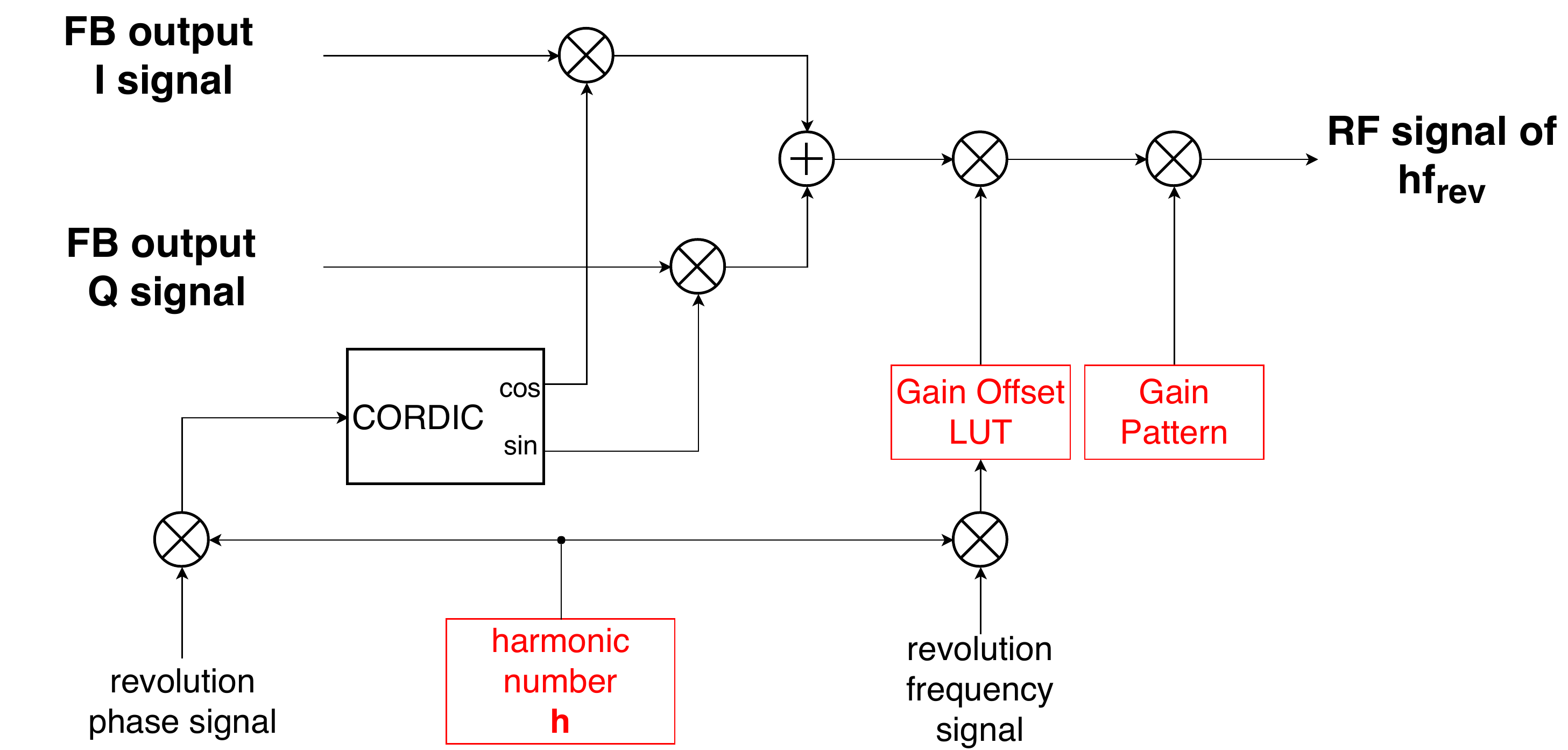}
\caption{Block diagram of the Digital Up Converter(DUC).}
\label{fig:DUC}
\end{figure}
\begin{figure}[b]
\includegraphics[width=\linewidth]{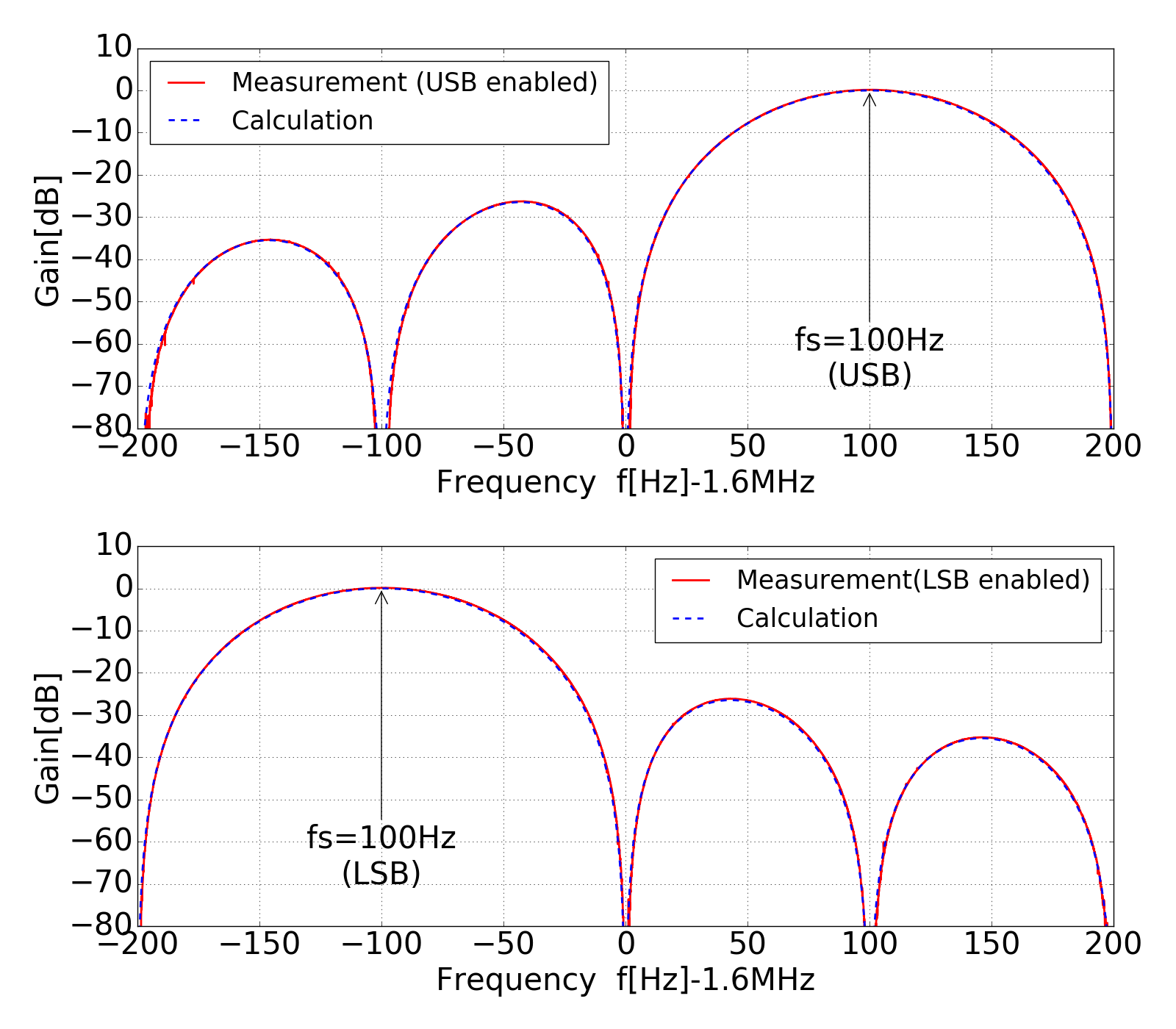}
\caption{The frequency response of the feedback processor with the synchrotron frequency set to be 100 Hz.}
\label{fig:freq_res_CIC_fs_track_100}
\end{figure}
\subsection{Digital Up Converter}

%\newpage
The feedback output is up-converted to a RF signal by the DUC.
Figure \ref{fig:DUC} shows the block diagram of the DUC.
The feedback output I/Q signal is mixed with the sine and the cosine signal of the harmonic frequency and summed together as a RF signal.
The gain adjustment is done after the up-conversion.
The amplitude frequency response of the RF cavity is compensated by the gain LUT addressed by the harmonic frequency.
An additional gain control can be done by a gain pattern. %set via EPICS.
The RF signals from all the harmonic blocks are summed and used as the input to the DAC.

\begin{figure}[b]
\includegraphics[width=\linewidth]{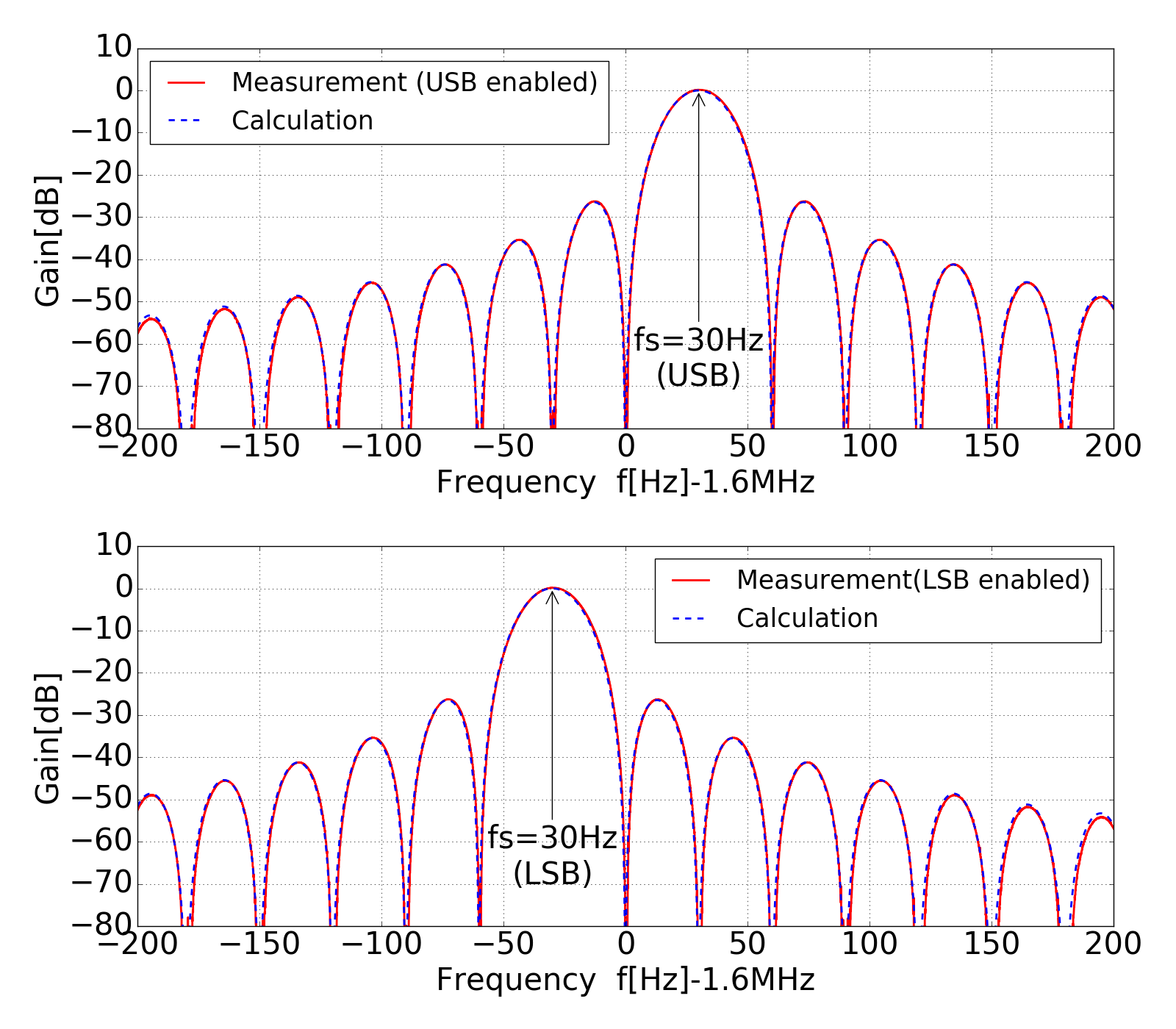}
\caption{The frequency response of the feedback processor with the synchrotron frequency set to be 30 Hz.}
\label{fig:freq_res_CIC_fs_track_30}
\end{figure}

\section{System Performance Test}
\subsection{Frequency response}

The frequency response of the feedback processor was measured with a network analyzer.
%
%\begin{figure}[htbp]
%\includegraphics[width=.9\linewidth]{NWA_setup.pdf}
%\caption{The configuration of the measurement of the frequency response.}
%\label{fig:fresponse_setup}
%\end{figure}
%
%Figure \ref{fig:fresponse_setup} shows the configuration of the measurement of the frequency response. 
The input and the output ports of the feedback processor were connected directly to those of the network analyzer.
%The input and the output of the feedback processor are connected to the network analyzer.
The revolution frequency pattern was set to 200 kHz with the harmonic number of 8.
The gain of the P controller was set to 1 and that of the I controller was set to 0 so that only proportional control was enabled. % to see the frequency response of the filter.
The reference I/Q pattern was set to $(I,Q)=(0,0)$ to see the frequency response of the filter.
The synchrotron frequency pattern was set to constant value for the measurement. % by the network analyzer.
Only one sideband was enabled for the output in the SSBF block to see the frequency response of the processor for each sideband.
%The frequency response of the CIC filter for the baseband harmonic detection was measured by bypassing the SSBF.

%Figure \ref{fig:freq_res_CIC_BB} shows the measured and calculated frequency response of the baseband CIC filter.
%\begin{figure}[htbp]
%\includegraphics[width=.6\linewidth]{GainVsFreq_ver613_zoom.png}
%%\includegraphics[width=.6\linewidth]{GainVsFreq_ver613_zoom.png}
%\caption{The frequency response of the CIC filter for the baseband harmonic detection.}
%\label{fig:freq_res_CIC_BB}
%\end{figure}
%The measured frequency response matched well with the calculation.
%The reduction of 40 dB above 186 Hz can be confirmed.

Figures \ref{fig:freq_res_CIC_fs_track_100} and  \ref{fig:freq_res_CIC_fs_track_30} show the frequency response of the feedback processor in the case of the synchrotron frequency at 100 Hz and 30 Hz,
respectively.
In both synchrotron frequency setup,
the system had maximum transmission with the gain around 0 dB at the selected sideband frequency,
and the DC component and the other sideband component were well suppressed below -80 dB.
The measured frequency response matched well with the calculation.
This proves that the SSBF worked as designed. % and the feedback system 

%The rejection of the DC component and the transmission of the synchrotron sideband moving along the frequency pattern were confirmed in two different synchrotron frequency setups.

%\begin{figure}[htbp]
%\includegraphics[width=\linewidth]{impedance.png}
%\caption{The frequency response of the fixed bandwidth CIC filter for the sideband detection.}
%\label{fig:freq_res_CIC_fs_fix}
%\end{figure}
%
%\begin{figure}[htbp]
%\includegraphics[width=\linewidth]{impedance.png}
%\caption{The frequency response of the high pass filter for the sideband detection.}
%\label{fig:freq_res_HPF}
%\end{figure}

%\newpage

%hoge hoge hoge hoge hoge hoge hoge hoge hoge hoge hoge hoge hoge hoge hoge hoge hoge hoge hoge hoge hoge hoge hoge hoge hoge hoge hoge hoge hoge hoge hoge hoge hoge hoge hoge hoge hoge hoge hoge hoge hoge hoge hoge hoge hoge hoge hoge hoge hoge hoge hoge hoge hoge hoge hoge hoge hoge hoge hoge hoge hoge hoge hoge hoge hoge hoge hoge hoge hoge hoge hoge hoge hoge hoge hoge hoge hoge hoge hoge hoge hoge hoge hoge hoge hoge hoge hoge hoge %hoge hoge hoge hoge hoge hoge hoge hoge hoge hoge hoge hoge hoge hoge hoge hoge hoge hoge hoge hoge hoge hoge hoge hoge hoge hoge hoge hoge hoge hoge hoge hoge hoge hoge hoge hoge hoge hoge hoge hoge hoge hoge hoge hoge hoge hoge hoge hoge hoge hoge hoge hoge hoge hoge hoge hoge hoge hoge hoge hoge hoge hoge hoge hoge hoge hoge hoge hoge hoge hoge hoge hoge hoge hoge hoge hoge hoge hoge hoge hoge hoge hoge hoge hoge hoge hoge hoge hoge hoge hoge hoge hoge hoge hoge hoge hoge hoge hoge hoge hoge hoge hoge hoge hoge 

\subsection{Measurement of the beam oscillation}

\begin{figure}[t]
\centering
\includegraphics[width=\linewidth]{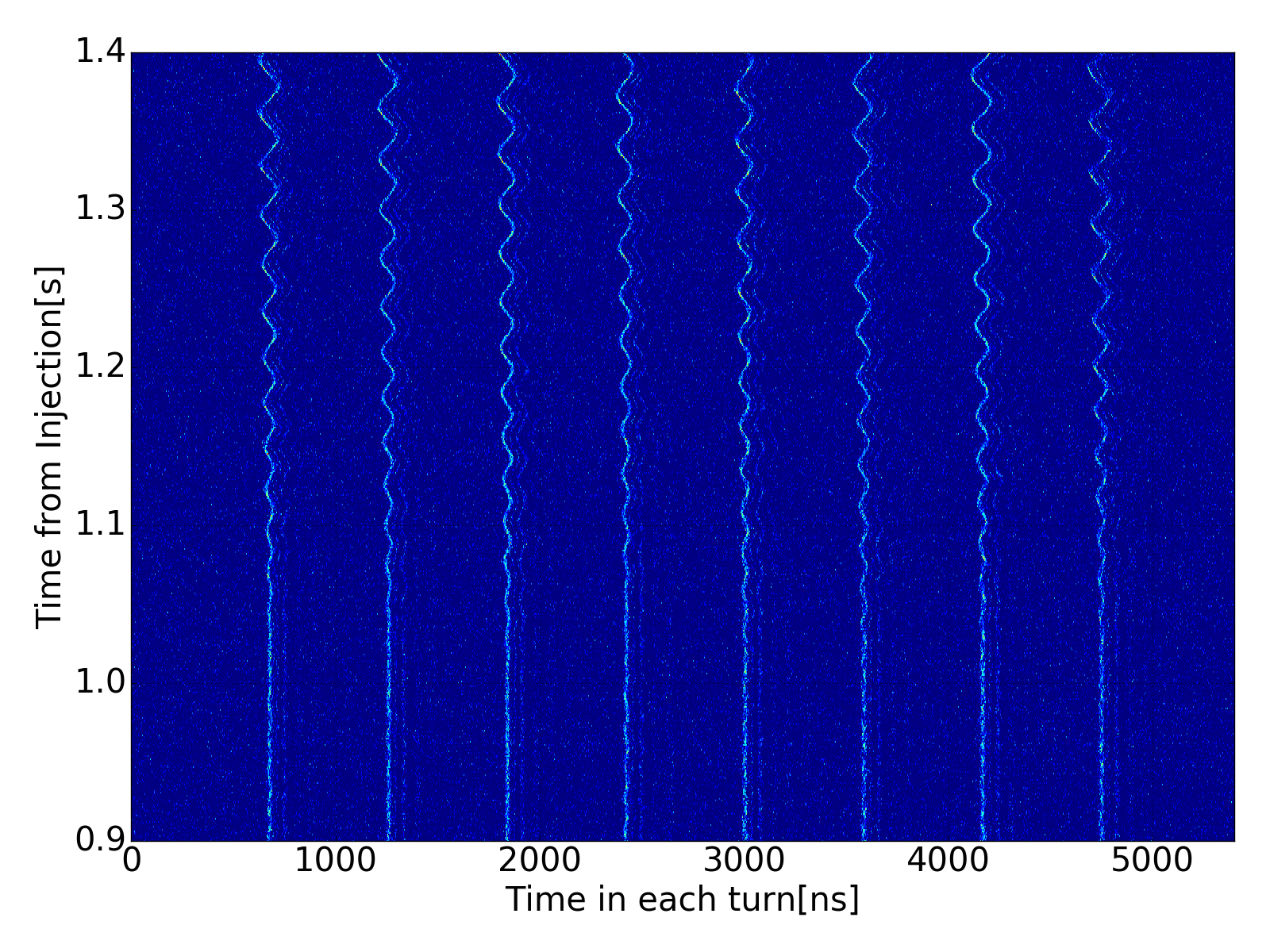}
\caption{The mountain plot for the beam excitation measurement.}
\label{fig:mountain_excite}
\end{figure}

The performance of the system to excite and detect the beam oscillation was measured with beam.
The measurement was done with the configuration for the beam operation which shown in the Fig. \ref{fig:FBsystem}.
The feedback loop was opened for the measurement.
The system kicked the beam with the excitation voltage pattern and monitored the beam oscillation.
The excitation voltage pattern was controlled by modulating the I/Q reference pattern.
The USB of the harmonic component of $h=8$ was excited by the modulation to have the CB oscillation of mode $n=8$. 
The calculated synchrotron frequency pattern was used for the sideband detection and also for the modulation in the feedback processor. 
A single RF cavity was used in the feedback system to generate the kick voltage of 2.5 kV.
%The kick voltage was set to 2.5 kV by using a single RF cavity.
The excitation was started from 1.0 sec after the beam injection.

The excitation measurement was done with 12-kW beam low enough to get the stable beam without any longitudinal oscillation.
Figure \ref{fig:mountain_excite} shows the mountain plot for the beam excitation measurement.
The beam oscillation growing from 1.0 sec can be seen in this plot.
This proves that the system can generate kick voltage large enough to control the beam oscillation.
%The growth in the oscillation also shows that the synchrotron frequency pattern matched with the beam.

\begin{figure}[t]
\centering
\includegraphics[width=\linewidth]{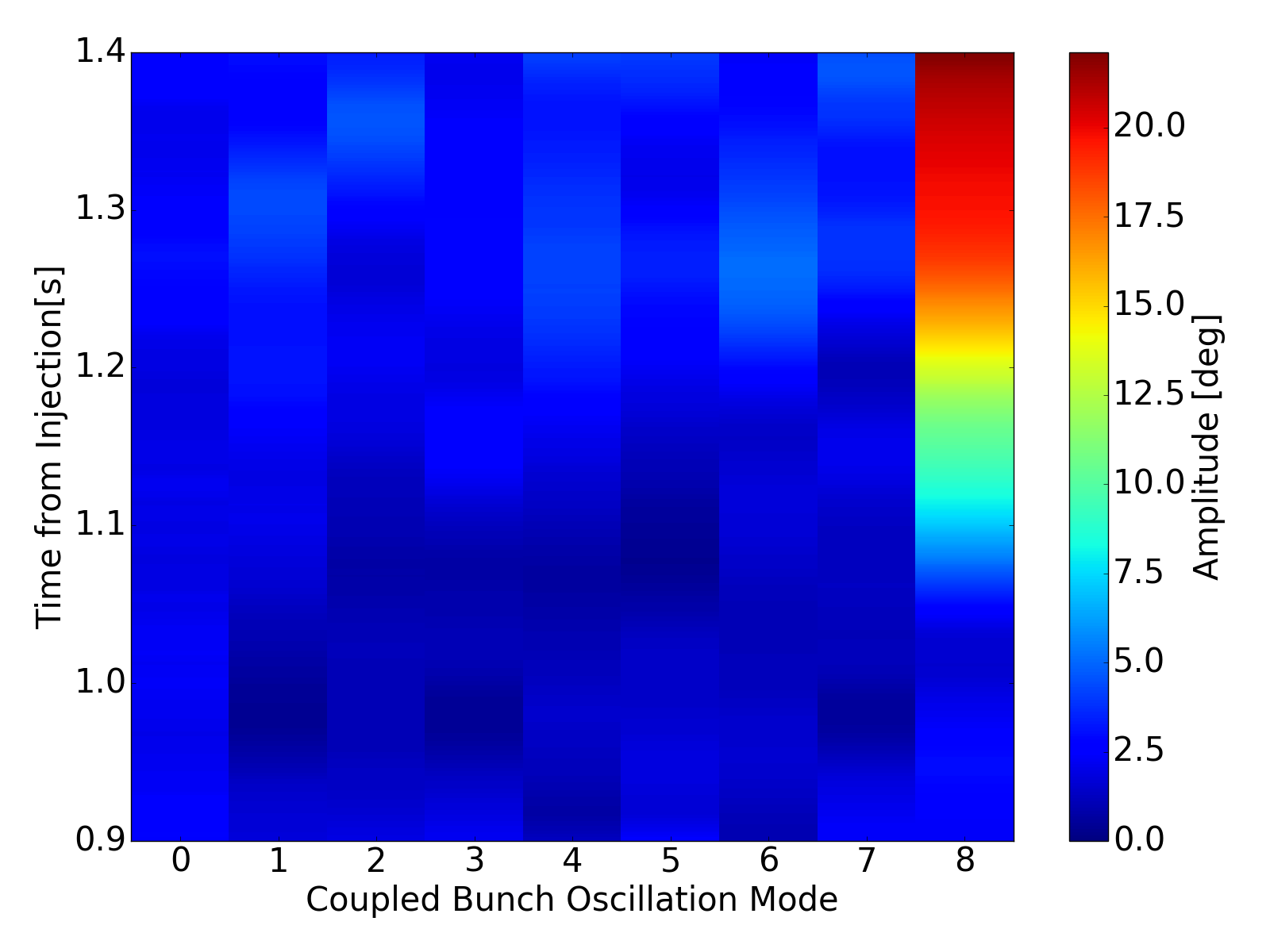}
\caption{The time variation of the oscillation amplitudes of each CB mode based on the bunch centers motion analysis.}
\label{fig:CBOana}
\end{figure}

The oscillation amplitude of each CB mode was obtained by analyzing the motion of bunch centers\cite{Damerau2007,Sugiyama:IBIC2017-TUPCF10}.
The motion of bunch centers were obtained from the analysis of the beam signal recorded by the oscilloscope.
Figure \ref{fig:CBOana} shows the variation of the oscillation amplitudes of each CB mode based on the bunch centers motion analysis.
No significant CB oscillation observed in all CB modes before the excitation,
and only mode $n=8$ has significant growth after the excitation.
This result shows that the CB oscillation of mode $n=8$ was successfully excited.
%
%\begin{figure*}[bt]
%\centering
%%\includegraphics[width=80mm]{CBO_SSBana_LSB_SSBData_1022_09.png}
%\includegraphics[width=.8\linewidth]{oscilloana.pdf}
%%\caption{The time variation of the amplitudes of the USBs of the harmonic components\cite{Sugiyama:IBIC2017-TUPCF10}.}
%%\label{fig:SSBresult_USB_excite}
%%\end{figure}
%%\begin{figure}[hbt]
%%\centering
%%\includegraphics[width=80mm]{CBO_SSBana_USB_SSBData_1022_09.png}
%\caption{The time variation of the amplitudes of the LSBs (left) and USBs (right) of the harmonic components in the oscilloscope analysis.}
%%\label{fig:SSBresult_LSB_excite}
%\label{fig:SSBresult_excite}
%\end{figure*}

%Figure \ref{fig:SSBresult_excite} shows the time variation of the amplitudes of the LSBs and USBs of the harmonic components in the offline sideband analysis of the oscilloscope data.
%The LSBs and USBs are extracted by using single sideband filtering technique\cite{Kriegbaum1977} in the offline analysis.
%The significant growth were observed only in the LSB of the $h=1,10$ harmonic component and the USB of the $h=8$ component.
%These sidebands correspond to the synchrotron sideband for the CB mode $n=8$.
%This result shows that the CB oscillation of mode $n=8$ was successfully excited.

\begin{figure*}[b]
\centering
\includegraphics[width=\linewidth]{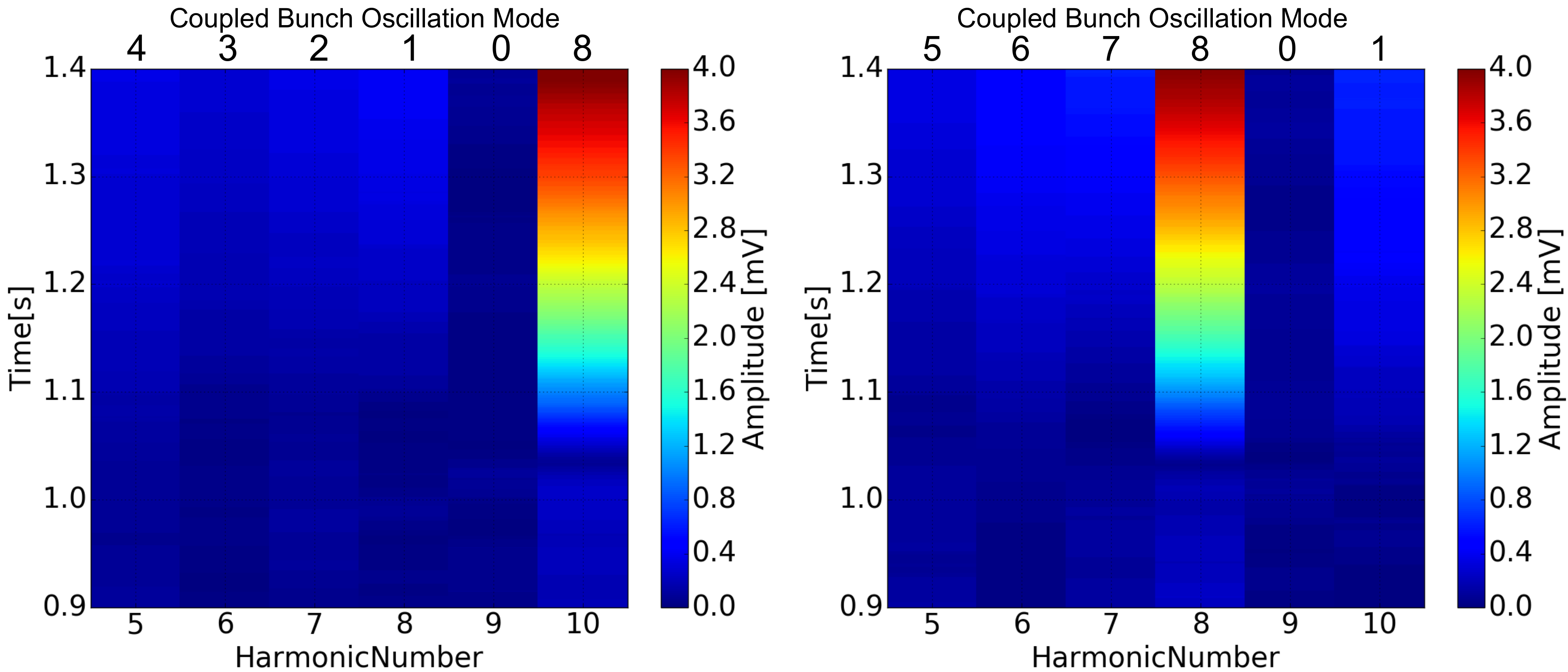}
%\caption{The time variation of the amplitudes of the USBs of the harmonic components\cite{Sugiyama:IBIC2017-TUPCF10}.}
%\label{fig:SSBresult_USB_excite_FB}
%\end{figure}
%\begin{figure}[hbt]
%\centering
%\includegraphics[width=80mm]{TUPCF10f9.png}
\caption{The time variation of the amplitudes of the the LSBs (left) and USBs (right) of the harmonic components detected by the feedback processor.}
%\label{fig:SSBresult_LSB_excite_FB}
\label{fig:SSBresult_excite_FB}
\end{figure*}

Figure \ref{fig:SSBresult_excite_FB} shows the time variation of the amplitudes of the the LSBs  and USBs  of the harmonic components detected by the feedback processor.
The amplitudes of the sidebands were obtained from the recorded waveform of the sideband I/Q complex signal at the output of the CIC filter in the SSBF.
The amplitude of the LSBs and USBs of the harmonic component with $h=5...10$ were recorded.
No significant oscillation was observed  in any sidebands of monitored harmonic component before the excitation,
and only the LSB of $h=10$ component and the USB of the $h=8$ component had significant growth with the excitation.
Both sidebands correspond to the CB mode $n=8$.
These sidebands  show the similar growth as those measured in the analysis of the oscilloscope data.
This result proves that the sideband detection function of the feedback processor worked as expected with the beam.

%\section{Discussion}

%hoge hoge hoge hoge hoge hoge hoge hoge hoge hoge hoge hoge hoge hoge hoge hoge hoge hoge hoge hoge hoge hoge hoge hoge hoge hoge hoge hoge hoge hoge hoge hoge hoge hoge hoge hoge hoge hoge hoge hoge hoge hoge hoge hoge hoge hoge hoge hoge hoge hoge hoge hoge hoge hoge hoge hoge hoge hoge hoge hoge hoge hoge hoge hoge hoge hoge hoge hoge hoge hoge hoge hoge hoge hoge hoge hoge hoge hoge hoge hoge hoge hoge hoge hoge hoge hoge hoge hoge hoge hoge hoge hoge hoge hoge hoge hoge hoge hoge hoge hoge hoge hoge hoge hoge 

\section{Summary and outlook}
We summarize the article as follows.

The longitudinal coupled bunch oscillation becomes serious in the J-PARC MR. 
To mitigate it for stable beam operation with higher beam intensities,
we developed a longitudinal mode-by-mode feedback system for the coupled bunch oscillation in the J-PARC MR.
The feedback processor detect and control the synchrotron sidebands of each CB mode utilizing the single sideband filter.

The frequency response of the filters in the feedback processor was measured with the network analyzer and was matched well with the calculation.
The sideband detection performance of the system was tested with exciting the 12-kW beam.
The CB oscillation of the selected mode was successfully excited and the oscillation amplitude of the coupled bunch oscillation measured by the system agreed with the oscilloscope analysis.

We will perform the beam test to suppress the beam oscillation by closing the feedback loop.
The measurement of the phase offset LUT with the beam in the feedback loop is a key to close the loop.
%The measurement of the transfer function of the system including the beam in the loop will be the key to close the loop.

\section*{Acknowledgments}
We would like to thank Heiko Damerau for fruitful discussions on the feedback for the CB instabilities.
%We would like to express our gratitude to the KEK Computer Research Center and the Network and Computing Service Center of Yamagata University for the computer and network resources.
We also would like to thank the Mitsubishi Electric TOKKI Systems Corporation  for their contribution including the hardware and the firmware development for the feedback processor. 
Finally, we would like to thank all members of the J-PARC accelerator group for their supports.
%This research was supported by KEK, 
%MEXT KAKENHI Grant Number 18071006,
%JSPS KAKENHI Grant Number 23224007,
%the Japan/US Cooperation Program,
%the DOE award DE-SC0002644 through a subcontract from the University of Michigan,
%and DOE award DE-SC0006497 to Arizona State University.
%


%\bibliographystyle{IEEEtran}
%\begin{thebibliography}{2}
%\bibitem{IEEEPDFRequirement401}
%IEEE Content Engineering, \emph{IEEE PDF Specification Version 4.10}. Available: http://www.ieee.org/documents/31296\_IEEE\_PDF\_Spec.zip.
%
%\end{thebibliography}
%
\bibliographystyle{IEEEtran}
\bibliography{IEEEabrv,ForPaper-RT2018.bib}
\end{document}